# Real-Time Monitoring and Driver Feedback to Promote Fuel Efficient Driving

Sandareka Wickramanayake[1], H.M.N Dilum Bandara[1,*] and Nishal A. Samarasekara[2]
[1]Dept. of Computer Science and Engineering, [2]Dept. of Transport and Logistics Management
University of Moratuwa, Sri Lanka
sandarekaw@cse.mrt.ac.lk, dilumb@cse.mrt.ac.lk, nishals@uom.lk

*Abstract* — Improving the fuel efficiency of vehicles is imperative to reduce costs and protect the environment. While the efficient engine and vehicle designs, as well as intelligent route planning, are well-known solutions to enhance the fuel efficiency, research has also demonstrated that the adoption of fuel-efficient driving behaviors could lead to further savings. In this work, we propose a novel framework to promote fuel-efficient driving behaviors through real-time automatic monitoring and driver feedback. In this framework, a random-forest based classification model developed using historical data to identifies fuel-inefficient driving behaviors. The classifier considers driver-dependent parameters such as speed and acceleration/deceleration pattern, as well as environmental parameters such as traffic, road topography, and weather to evaluate the fuel efficiency of one-minute driving events. When an inefficient driving action is detected, a fuzzy logic inference system is used to determine what the driver should do to maintain fuel-efficient driving behavior. The decided action is then conveyed to the driver via a smartphone in a non-intrusive manner. Using a dataset from a long-distance bus, we demonstrate that the proposed classification model yields an accuracy of 85.2% while increasing the fuel efficiency up to 16.4%.

* Corresponding author

# I. INTRODUCTION

The rising number of motor vehicles demand several forms of energy sources. Although electric and hybrid vehicles are becoming popular, the world is still dependent mostly on fossil fuel sources. For example, in Sri Lanka, 82.3% of the total fossil fuel consumption is attributed to the transportation sector (Sri Lanka Sustainable Energy Authority, 2014) whereas in the United States 92.2% of the total energy requirement for transportation is fulfilled by fossil fuels (U.S. Energy Information Administration, 2016). Hence, it is imperative to improve the fuel efficiency of vehicles to reduce fuel usage and thereby save money. Especially when it comes to the fleet industry, whose central expenditure is the fuel cost, millions could be saved annually by adhering to efficient fuel consumption. Besides, saving fuel means saving the environment.

Various factors contribute to the fuel consumption of a vehicle. Demir, Bekta'ş, & Laporte (2014) classified the factors influencing fuel consumption into five categories as vehicle, environment, traffic, driver, and operations. They concluded that while most of the fuel consumption models pay attention to the impact due to the vehicle, environment, and traffic, little consideration is given on the impact due to driver behavior and operation. While efficient engine/vehicle design and intelligent route planning can achieve substantial fuel savings, further fuel savings could be achieved by optimized driving. For example, Gonder, Earleywine, and Sparks (2012) showed that efficient driver behavior could provide up to 20% fuel savings. While the drivers could be educated on general guidelines to save fuel, more savings could be achieved through individual feedback. Such feedback is typically based on historical data, which is used for driver training and appraisals. While such training could improve efficiency over time, it may saturate at a suboptimal level. However, we believe that much more useful feedback can be given through real-time driver monitoring, where we could assist the drivers to change their driving behaviors while on the road to maintain an efficient and safe driving behavior. The results could be immediate and more significant, as the driver is carefully pushed to reach a more optimum efficiency level. However, it is essential to monitor and provide feedback in a non-intrusive manner while being aware of the environmental conditions, being conscious of the impact to the other traffic, and encouraging safe driving.

While analyzing the impact of driver behavior on fuel usage, the related work mostly considers only the driver dependent parameters. Nevertheless, external factors such as weather, road traffic, road topography, and road conditions also influence the driving behavior to a greater extent. To provide more useful feedback, one should consider both the driver-dependent and driver-independent factors influencing fuel usage. Such driver behavior and fuel consumption analysis typically require high-resolution data from GPS-based tracking devices, various sensors, and other external sources. Given the volume, diversity, and uncertainty of data, sophisticated data mining, and big data analytics techniques are required to identify fuel-inefficient driver behaviors and to provide useful recommendations for individual drivers in real time.

In this paper, we propose a novel framework to encourage drivers to follow fuel-efficient driving behaviors. A random-forest based classification model is developed using historical data to evaluate the driving behaviors for fuel efficacy considering vehicular, GPS, weather, and traffic data collected from sensors on the vehicle, as well as from external data sources. If a driving behavior is detected to be inefficient, the system attempts to infer the possible reason(s) behind inefficiency and what action the driver should perform to bring the vehicle back to a fuel-efficient state. For this, we propose to use a fuzzy-logic inference system. The suggested action is then conveyed to the driver via a mobile app as voice commands. A mobile app is selected to provide driver feedback due to the pervasiveness and to enhance usability and flexibility, as well as to reduce costs. Voice commands are used, as they are non-intrusive and would not compromise safety (Hammerschmidt & Hermann, 2017).

The critical step in this solution is the identification of driving events (i.e., periods of driving), which are fuel inefficient. While a classification model could be used, it is nontrivial to label the fuel efficiency of a given driving event considering all driver-dependent and driver-independent parameters, as well as the sheer volume of data. We address this problem by automatically clustering data points in a high-dimensional space, and then manually analyzing those clusters for the fuel efficiency. We then apply labels to the clusters, and thereby individual data points are labeled. This labeling is more accurate than labeling just by looking at the fuel usage, as the fuel usage depends on external conditions where fuel-inefficient driving behaviors might be inevitable under some conditions. For example, the driver might be forced to drive slowly due to heavy rain, snow, or excessive idling in a traffic jam.

We demonstrate the proposed technique using a dataset from a long-distance bus. This dataset provides an ideal test bench, as it includes all types of external conditions such as driving in city, suburban, and rural areas; driving within peak and off-peak hours, and night driving; and driving through both flat and mountainous regions. As our framework does not use any parameter that is specific to this bus (e.g., load) or its route (e.g., latitude or longitude), it could be generalized to other cases and vehicles. We selected hierarchical clustering to cluster the data points based on their attributes, and then a random-forest based model is used to classify the clusters as fuel efficient or inefficient. The developed classification model has an accuracy of 85.2%. To simulate the benefits of the proposed mechanism, we compared the fuel economy of a given journey with the historically best fuel economy for the same location, time, altitude, and weather depending on the suggested action of the fuzzy-logic inference system. Comparing with the best historical case, potential savings due to driver feedback could be up to 16.4%.

The rest of the paper is organized as follows. Section II presents the related work and explains the importance of combining both driver dependent and independent parameters. A high-level overview of the proposed solution is presented in Section III. Section IV presents the proposed methodology, the dataset, the clustering procedure, the classification model, and how driver action items are determined. Evaluation of the proposed solution is presented in Section V while concluding remarks are given in Section VI.

## II. RELATED WORK

Walnum and Simonsen (2015) analyzed the impact of driving behaviors on fuel economy based on a dataset of heavy-duty trucks traveling in Norway. The dataset consists of parameters such as the weight of payload, route, brake horsepower, average speed, cruise-control use, running in the idle state, driving in highest gear, brake applications, number of stops, and a dummy variable for seasonal variation. Authors concluded that while road and vehicle conditions have a higher impact under certain conditions, in general, driver behavior significantly affects the fuel economy.

FleetCarma (n.d), a clean-tech company, attempted to quantify the effect of driver behavior on fuel consumption. FleetCarma introduced a metric named *Eco-score* to measure driver aggressiveness. Through the analysis of real-world driver behavior, they demonstrated that the impact of driving behavior on fuel economy varies with the mass of the vehicle, greater the mass, higher the fuel consumption sensitivity to the driving behavior. FleetCarma also showed that driver behavior could be enhanced through consistent feedback, and thus fuel economy of the vehicle can be improved.

Further, the impact of individual driver feedback on eco-driving has been well-studied (Toledo & Shiftan, 2016; Rolim, Baptista, Duarte, Farias, & Pereira, 2017a; Ayyildiz, Cavallaro, Nocera, & Willenbrock, 2017; Rolim, Baptista, Duarte, Farias, & Pereira, 2017b) In some work feedback is given real time (Rolim, Baptista, Duarte, Farias, & Pereira, 2017a; Rolim, Baptista, Duarte, Farias, & Pereira, 2017b). In others feedback is given off-line, e.g., as a summary report at the end of the journey, end of the day or for every two weeks (Toledo & Shiftan, 2016). All these studies advocate the fact that individual feedback contributes to enhancing the fuel efficiency of the vehicle than traditional group-based training programs.

Analysis of driver data using statistical models and clustering techniques is presented by CGI (2014). The authors suggest clustering driver data into four clusters based on trip distance and average speed and model fuel consumption for each cluster separately. They have used cruise control, idling, hard acceleration, harsh break, high RPM, and rollout to rank driver behavior. Authors propose to provide driver feedback daily and found that 10% - 30% of the variation in fuel consumption can be attributed to driver behavior. However, only weather information has been considered as an influential external factor. Even under weather information, only the average temperature and wind speed have been considered. However, other weather conditions like rain, fog, hail, and snow also are known to have a more significant impact.

Linda and Manic (2012) proposed an *Intelligent Driver System* to improve fuel economy by learning fuel-efficient driving behaviors. The proposed intelligent driver system gradually builds a model of historically most fuel-efficient driving behavior for a fixed set of routes. This system considers both vehicle performance and GPS data while modeling driver behavior. While driving, the velocity of the vehicle is compared with the calculated optimal velocity for that specific location. Then a fuzzy Proportional-Derivative (PD) controller is used to determine the best control action to take the vehicle to the optimal velocity. One drawback of this solution is its inability to use beyond the predefined set of routes. Furthermore, the solution does not consider the impact due to weather and traffic conditions, which tend to change with time of the day, week, and year. Hence, sometimes the control action determined by this system would not be the optimal action, and it would be unsafe to adhere to those instructions. For example, even though historically optimal speed for a route in a given area is 70 $kmh^{-1}$, under heavy rain, it would be inappropriate to suggest driving at that speed.

Gilman et al. (2015) developed a reference architecture for context-aware driver assistance systems to provide personalized assistance for fuel-efficient driving. The prototype collects, integrates, and analyses different data such as vehicle parameters, weather, and traffic, and then provide off-line driver feedback to improve future driving behaviors. While the analysis of multiple factors gives an enhanced model, it could be more effective if the feedback can be given in real time.

The study of existing work in driver monitoring and feedback revealed that current frameworks suffer from one or more of the following drawbacks:

- Feedback is given off-line, for a trip or for one day of traveling. It prevents drivers from understanding their mistakes on the road and adopting a fuel-efficient driving behavior immediately.
- Feedback is given manually. Manual feedback is costly as it needs domain expertise and would not be plausible to provide in real time.
- Select a limited set of relevant attributes that impact the fuel consumption

Hence, there is a need to build a model that is comprehensive in capturing both the driver dependent and independent parameters, could be used for real-time detection and feedback, as well as beyond predefined routes.

III. SYSTEM OVERVIEW

To enhance the fuel economy of fleet vehicles, we propose a framework that monitors the drivers and provides individual feedback in real time. This system takes both the driver dependent and driver-independent influences into consideration in deciding whether a particular driving behavior is fuel efficient or not. Figure 1 illustrates a high-level overview of the proposed system. We assume that the vehicles are equipped with a GPS-based tracking system and a high-precision fuel sensor. Conventional floater-based fuel sensors are over-sensitive to potholes and bumpers on the road, upward and downward slopes, and rapid acceleration and deceleration. Therefore, high-precision fuel sensors are typically used to obtain more accurate fuel levels. Vehicle-related data such as the speed, acceleration, current location of the vehicle, and fuel level are pushed to the cloud-based backend in near real time. This could be archived using commodity technologies such as 3G/4G or Machine-to-Machine (M2M) communication.

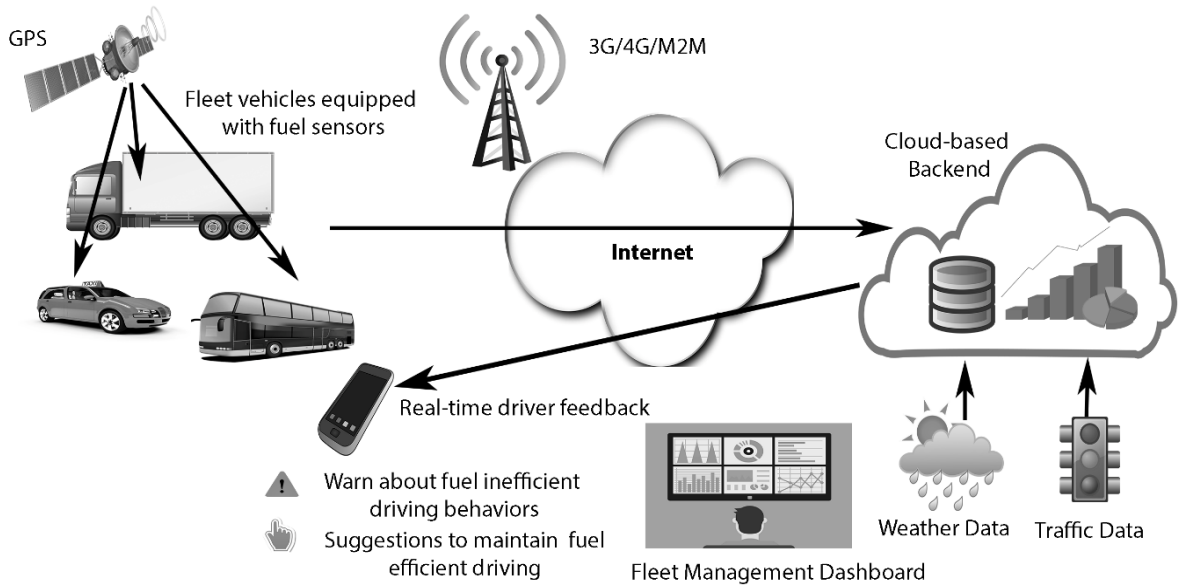

Figure 1: System overview.

An analytical engine running in the cloud-based backend combines GPS and fuel data, as well as other relevant data such as weather and traffic conditions to evaluate the current driving behavior. Relevant weather and traffic data could be pulled from third-party data sources using a REST API. If the driving behavior is determined to be ineffective, the analytical engine determines a suitable corrective action to take the vehicle back to the fuel-efficient state. The action is then communicated to the driver as a voice alert using a mobile app.

Further, vehicle owners and fleet managers could be given a dashboard, which summarizes driving behavior and its impact on fuel consumption (see Figure 1). The dashboard may also indicate other metrics such as the percentage of time a particular driver adheres to fuel-efficient driving habits and causes for inefficiency. Such a dashboard is useful in driver coaching, as well as to appraise the performance of drivers regarding driving behavior and fuel efficiency.

Figure 2 illustrates the flow diagram of the analytical engine residing in the cloud-based backend. Once GPS, fuel consumption, and driving behavior related data arrive at the system, data are preprocessed. Preprocessing is required to clean the data, as well as to derive new parameters such as *isIdling* (i.e., whether the vehicle is idling), *hour* (i.e., time of day), and fuel mileage. *Fuel mileage* is calculated as the ratio of distance traveled in *one minute* to the fuel consumption of the vehicle in that minute. Vehicular data and weather data are then fed into a classification model, which determines the fuel efficiency of the driving event. A *driving event* is defined as the information related to the driving behavior within one-minute, such as the average speed, average acceleration, elevation change, whether the vehicle is idling, and weather condition within that one minute. If a given driver's behavior is identified as fuel-inefficient, then it is checked for idling. If the inefficiency has been caused by excessive idling, the driver is advised to stop the engine. If not, data is fed into a fuzzy logic inference system to determine the control action. The predicted control action is then transferred back to the driver as feedback. This process continues per every minute as data about new driving events keep arriving from the vehicle. We selected one minute as the duration of a driving event based on several factors. If we reduce the period, e.g., 30-sec, the system would not be able to differentiate the vehicle being idling and vehicle stopping at a traffic light. Furthermore, drivers would find it to be too intrusive to receive feedback for every 30-sec and be discouraged to use the system. In contrast, aggregating driving behavior data for a more extended period, such as 5-min could lead to information loss. For instance, average speed may conceal a mix of lower speeds and higher speeds that result in lower fuel efficiency compared to maintaining a steady speed. Moreover, FleetCarma (n.d) has emphasized through their analysis that the highest fuel saving is possible by removing 1-min idle events. According to their findings, fleets could save $583.72 yearly by eliminating 1-min idle events, whereas elimination of 5-min idle events could only save up to $ 503.23 per year.

IV. METHODOLOGY

*A. Dataset and Parameter Selection*

The dataset corresponds to a long distance, public bus in Sri Lanka. The bus starts from the Depot around 4:00 pm and then goes to Colombo (i.e., the commercial capital) along the A2 road. Then the bus leaves Colombo central bus stand at 7:00 pm and travels along the A4 and AB10 roads and reaches the destination around 7:00 am on the following day. Altogether, the bus travels ~365 km in one direction. The return journey is along the same route and typically between 4:00 pm to 7:00 am on the following morning. About one-third of the journey is through a mountainous region. This particular route captures almost all the external conditions a vehicle could encounter in real-world driving. For example, the bus goes through urban, suburban, rural, and mountainous areas, as well as driving times, including peak, off-peak, and night driving.

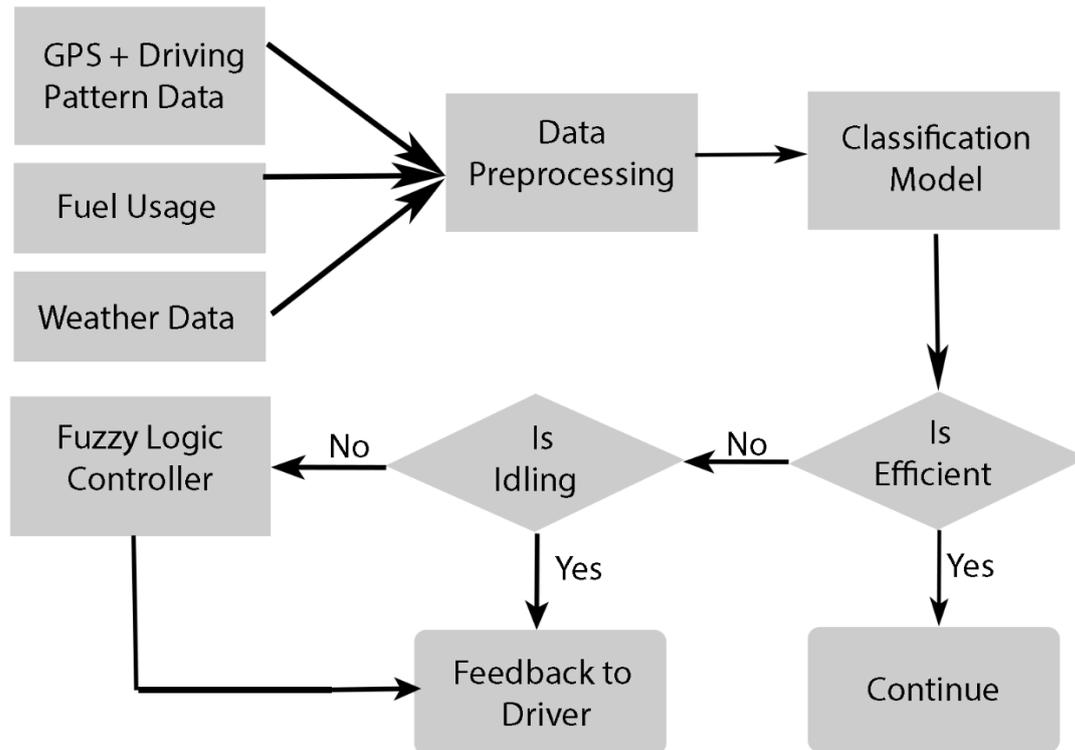

Figure 2: Data flow for one driving event.

The bus is fitted with a GPS-based tracking device and a capacitive, high-precision fuel sensor. Collected data is pushed to a cloud server in near real time using a GPRS module. The dataset consists of outward and inward journeys between May 13 and August 31, 2015. The dataset contains the following parameters:

- Timestamp (date and time)
- Longitude (Min: 5.9186°N, Max: 9.8355 °N)
- Latitude (Min: 79.5166 °E, Max: 81.8791 °E)
- Bearing (00 to 3600)
- Elevation (Min: 0m, Max: 2,524m)
- Distance traveled (km) – the distance bus traveled since the last record
- Speed ($kmh^{-1}$)
- Acceleration ($kmh^{-2}$)
- Ignition status (1 – Ignition On, 0 – Ignition Off)
- Current battery voltage (Min: 0v, Max: 29v)
- Fuel level (Min: 0L, Max: 218L)
- Fuel consumption (L)

The GPS-based tracking device sends data to the cloud when the bus takes a turn or every 17 seconds (whichever occurs first). Other events of interest, such as rapid acceleration, deceleration, and ignition on/off state, are also tracked. However, they are time-stamped with the last time reading from the GPS (due to hardware constraints time is read from the GPS module once every 5 seconds). This leads to uneven sampling. Therefore, to make the sampling rate consistent, the dataset was aggregated to one minute. Among the available parameters, we selected speed and acceleration as the driver dependent parameters, and elevation change and time of day as the environmental parameters. Fuel consumption is chosen as the dependent parameter. Fuel consumption is typically defined as either Kilometer per Liter (km/L) or Liters per 100 km (L/100km). A detailed exploratory data analysis of the dataset is presented in Wickramanayake and Bandara (2016). To capture the impact of road traffic, we considered the *time* of day. Time of day was derived by rounding off the GPS timestamp to the nearest hour (e.g., 12:00, 13:00, 14:00 and so on). Change of elevation captures the changes in road topography. A parameter to detect whether the vehicle is idling excessively (*isIdling*) was also derived from the data. We considered the vehicle to be excessively idling, if the speed is zero for more than one minute, while engine ignition status is on. Another essential but indirect environmental factor that affects fuel efficiency is the weather condition. When the weather is bad, drivers are forced to slow-down, which results in lower fuel efficiency. The weather condition was obtained using a REST API provided by the Historical or past weather API (n.d) of the World Weather Online (WWO API) developer portal. WWO API provides a detailed weather report for a given location, date, and time. However, for our analysis, we only extracted the weather descriptor (i.e., weather condition) for a given location, date, and time based on the bus's route and schedule. Weather descriptors provided by WWO API include sunny, clear, partly cloudy, cloudy, overcast, patchy rain nearby, light drizzle, light rain shower, moderate rain, moderate or heavy rain, mist, and fog.

### B. Clustering driving events

We aim to classify driver behaviors as fuel efficient or inefficient. However, it is essential to recognize that sometimes the drivers will not be able to follow fuel-efficient driving behaviors due to external factors such as traffic and weather conditions. Hence, we cannot solely rely on fuel consumption to label driving behaviors as either fuel efficient or inefficient. For instance, a driver is compelled to drive slowly in heavy rain to ensure safe driving. Consequently, the overall fuel consumption will increase, leading to lower efficiency. If we only focus on fuel efficiency, the driving events during the rain would be labeled as inefficient. While the natural response is suggesting that the driver should speed up, it is neither practical nor safe. Thus, we must consider relevant external factors in labeling our dataset.

Manually tagging individual data points as efficient or inefficient driving, considering all the influencing parameters and a large volume of data is a tedious task. Instead, we propose to use an unsupervised clustering technique to cluster the data points into different clusters in a high-dimensional space. Then we take the assistance of a domain expert(s) to analyze those clusters and label them as either fuel efficient or fuel inefficient considering not only the fuel efficiency, speed, and acceleration, but also weather, road condition, and traffic. Once the clusters are labeled, respective data points could also be labeled based on their cluster membership.

To understand the clustering problem at hand, let us consider the following simplified clustering task. We combined two significantly different set of data points along the bus route; data from a two-hour (17:00 - 19:00) drive close to Colombo (an urban area) and a two-hour (22:30 - 00:30) drive close to Udawalawa (a rural area). Figure 3 shows a scatter plot of the selected data points, where the fuel consumption is plotted against longitude. Circles indicate the data points close to Colombo, while triangles indicate the data points close to Udawalawa. We expect the chosen clustering algorithm to identify at least two clusters, even without geographical information. A careful analysis of the distribution of data points reveals the following characteristics of this clustering problem:

- The number of clusters is unpredictable for a given journey. As it depends on external factors, the number of clusters would vary widely.

- Clusters are not spherical.
- Clusters have uneven sizes. (i.e., different number of cluster members).
- Clusters have different densities, e.g., in Figure 3, the small cluster around $79.8^0$ has a higher density than other clusters.
- Clusters do not adhere to a normal distribution.

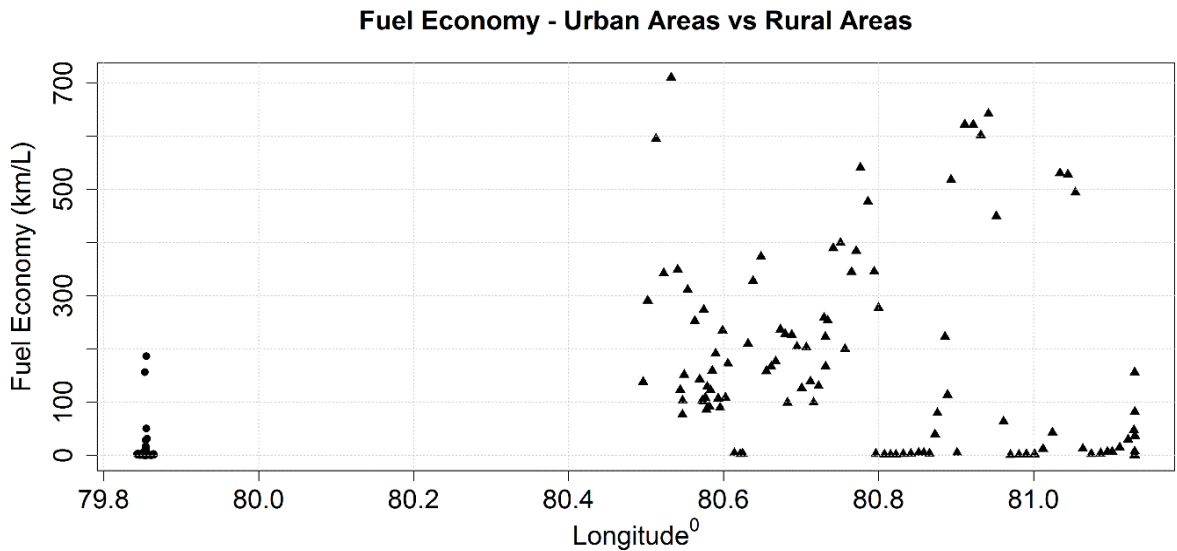

Figure 3: Fuel usage in urban and rural areas. Circles – urban area and Triangles – rural area.

According to a survey on different clustering techniques carried out by Berkhin (2006), agglomerative clustering is more suitable to cluster data points with the above characteristics. Berkhin has identified that the flexibility of exploring at different levels of granularity, ability to handle any form of similarity or distance metric, and applicability to any attribute type as the key advantages of hierarchical agglomerative clustering techniques. Further, if upper-level clusters on the dendrogram are not providing enough information to label the clusters, we can drill down to lower levels in the hierarchy, and then get smaller clusters and analyze their properties. These properties of hierarchical clustering enable more accurate labeling of fuel efficient and inefficient events. Thus, we choose hierarchical agglomerative clustering to cluster driving events.

To implement agglomerative hierarchical clustering "hclust" function in R was used. We used Euclidean distance to measure the distance between data points. While other distance metrics are supported by R, such as maximum, Manhattan, Canberra, binary, and Minkowski were tried, Euclidean distance resulted in better clusters. In clustering fuel data, *better clustering* refers to the ability to cluster events with the same external conditions into the same cluster, specifically at the lower levels of the hierarchy. "hclust" function (R Documentation, 2016) supports several algorithms such as ward.D2, single, complete, average, mcquitty, median, and centroid for agglomeration. After trying out those algorithms, we identified that ward's algorithm (i.e., ward.2D) gives better clusters, as it is less susceptible to noise and outliers.

The dendrogram in Figure 4 shows the cluster hierarchy for the dataset considered in Figure 3. We considered speed, acceleration, *isIdling*, elevation change, time of the day, and weather condition as the attributes for clustering. The location of the vehicle was not considered as an attribute, though it is used in Figure 3 and 5 for illustration purposes. As seen in Figure 4, cutting the tree at level four results in seven clusters. Figure 5 plots the same data points, labeled according to the clusters they belong to (different symbols indicate different clusters).

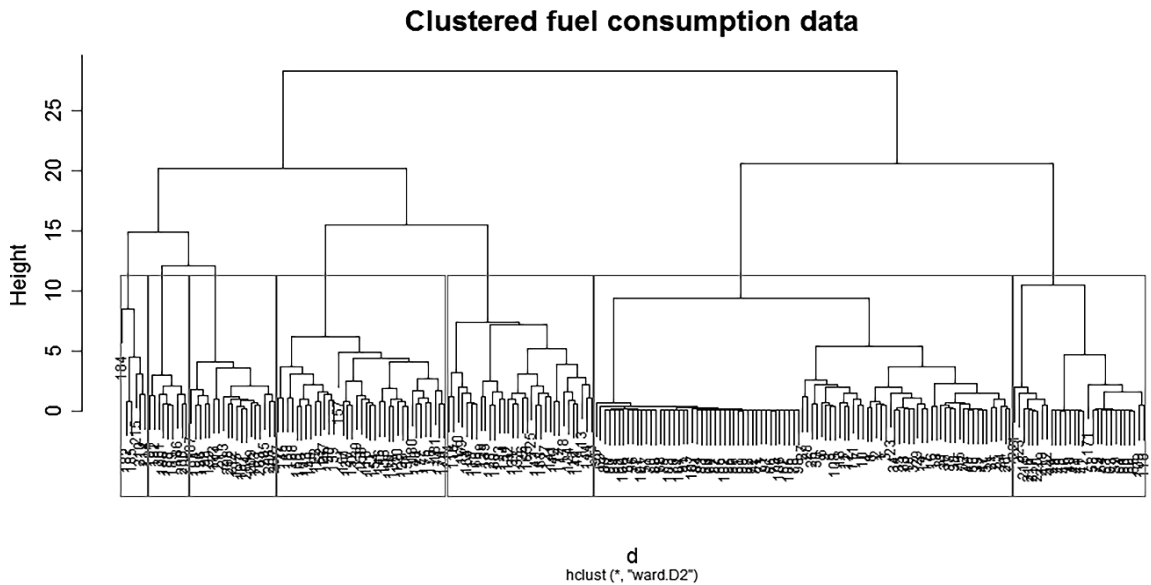

Figure 4: Dendrogram of clusters produced by hierarchical clustering.

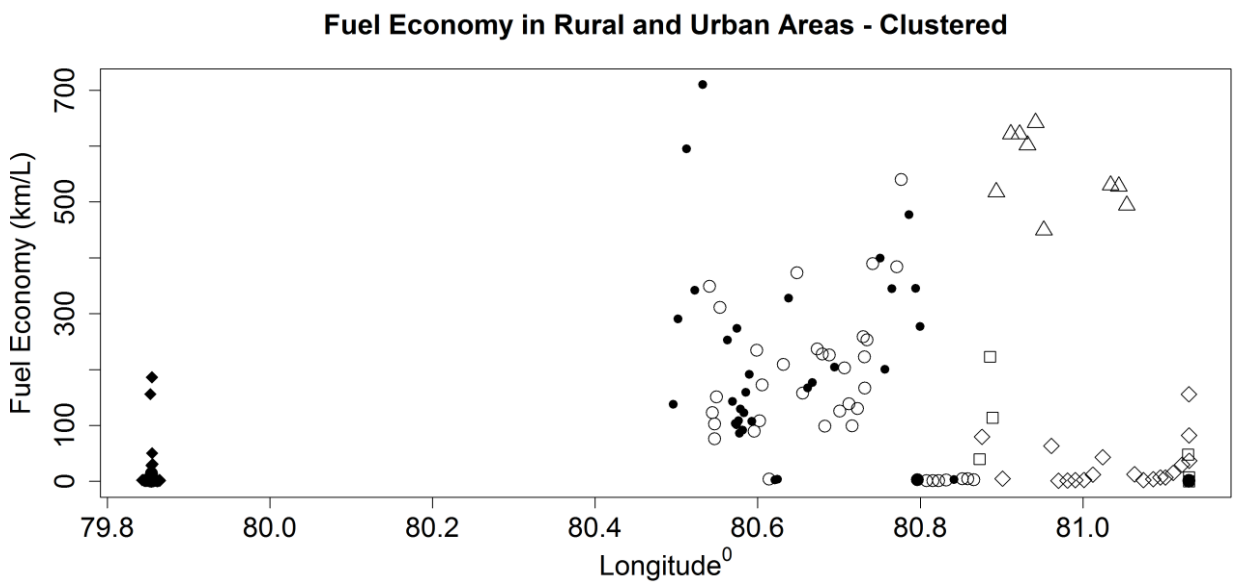

Figure 5: Fuel consumption in urban and rural areas after assigning to seven clusters.

Table I summarizes the set of identified clusters. The summary includes mean speed, acceleration, and elevation change of the cluster, whether most of the driving events in the cluster are idling events, most probable time of the day, weather condition, and mean fuel economy of the cluster. Since a driving event is the driver's behavior for a period of one-minute, very-high fuel economies are expected when the bus is cruising at its optimum or near optimum speed due to almost zero-fuel consumption. In labeling clusters, one can directly label cluster two as inefficient, not only because of the lower mean fuel usage (5.28 km/L) but also due to excessive idling at midnight. This is probably because the long-distance bus has stopped for a tea break, while the engine is running. If we consider the first cluster, both the fuel consumption (11.44 km/L) and mean speed (6.86 kmh$^{-1}$) are relatively low. However, the time of the day is 17.00, a peak traffic hour and the bus is running through Colombo, the capital of Sri Lanka. Therefore, one can attribute this lower speed to road traffic. Asking the driver to speed up the vehicle under this condition is not practical. This provides good evidence to prove the argument that a driver cannot be accounted for inefficient fuel usage just based on mean fuel usage. Cluster six (see Table I) is a contrasting case. Even though all external conditions are favorable, the mean fuel usage is not acceptable. Thus, cluster six was labeled as inefficient. In the same way, all the resulting clusters were analyzed by domain experts and labeled for their fuel efficiency, considering all the driver dependent and external factors.

Table I: Summary of each cluster derived using hierarchical clustering.

| Cluster No | Mean Speed (kmh$^{-1}$) | Mean Acceleration (kmh$^{-2}$) | Mean Elevation Change (m) | Is Idling (Mode) | Time of the day (Mode) | Weather Condition (mode) | Mean Fuel Economy (kmL$^{-1}$) (Distance in 1 min/ Fuel Consumption in 1 min) | Fuel Efficiency |
|---|---|---|---|---|---|---|---|---|
| 1 | 6.86 | -14.56 | -0.020 | 0 | 17.00 | Clear | 11.44 | Efficient |
| 2 | 0 | 0 | 0 | 1 | 00.00 | Partly Cloudy | 5.28 | Inefficient |
| 3 | 45.89 | -5.53 | 6.360 | 0 | 23.00 | Mist | 214.86 | Efficient |
| 4 | 45.99 | -109.83 | -7.379 | 0 | 23.00 | Mist | 167.25 | Efficient |
| 5 | 28.35 | -6,818.00 | 5.025 | 0 | 00.00 | Mist | 71.88 | Efficient |
| 6 | 61.12 | 273.00 | -0.960 | 0 | 00.00 | Partly Cloudy | 29.57 | Inefficient |
| 7 | 62.77 | 252.54 | -0.050 | 0 | 00.00 | Partly Cloudy | 556.30 | Efficient |

In labeling the whole dataset, each journey was clustered separately because different journeys might have driven by different drivers. While it is known that several drivers drove the bus on different days, the dataset does not contain information on who drove on a given day. Clustering each journey separately would eliminate a specific driver's behavioral impact on fuel consumption.

C. *Classification of Fuel Usage*

Once the historical data points are labeled, the next step is to develop a classification model that would classify a new data point as either fuel efficient or inefficient. The dataset we are analyzing is not linearly separable and known to have outliers. Therefore, the random forest was selected as the classification technique, as it could handle non-linear features, high-dimensional data, and many training samples.

Random forest is a predictive ensemble model based on a collection of decision trees (Rokach, 2010). Instead of making the prediction based on only one tree, random forest uses a set of trees to make the decision. Like other bagging models, the random forest also constructs each decision/regression tree using a bootstrap of sample data. However, the tree building procedure is different from others (Liaw and Wiener, 2002). Instead of splitting trees using the best split among all variables, in a random forest algorithm, each node is split using the best among a subset of predictors randomly chosen at that node. This enables random forest to be robust against overfitting and be outstanding among many other classifiers, including discriminant analysis, support vector machines, and neural networks. Moreover, Herrera et al.

(2010) demonstrated that random forest is reasonably fast to build and can be easily parallelized.

We used the random forest algorithm in R *random forest* package to build the classification model. *mtry*, i.e., the number of variables randomly sampled for a split, was set to three as it gave the least Out-Of-Bag (OOB) error estimate of 14.13%. *ntree*, i.e., the number of trees within the ensembles was set to 500 based on the empirical evidence. Historical data that were labeled via clustering in the previous step was used to training the random-forest based classification model.

D. *Determining the Control Action*

When the classification model detects a particular driving behavior as fuel inefficient, the next task is to determine what the diver should do to bring the vehicle back to a fuel-efficient state. As seen in Figure 2, the decision-making process has two steps. First, the proposed system checks whether the inefficiency is due to excessive idling. If so, the system sends the feedback suggesting the driver stops the engine. If idling is not the reason for the detected inefficiency, the system then uses a fuzzy-logic inference system to determine the control action. Fuzzy-logic inference systems have been successfully used in many applications in the transportation domain by Linda and Manic (2012) and Aljaafreh, Alshabatat, and Al-Din (2012). The reason for the popularity of Fuzzy-Logic Controllers (FLC) is their ability to model real-world ambiguous reasoning. FLC emulates the expert knowledge in the form of linguistic rules (Berenji & Khedkar, 1992). We choose driver-dependent influences on fuel usage, speed, and acceleration as inputs to the fuzzy logic system. Figures 6 and 7 show the corresponding membership functions, respectively. According to the exploratory analysis of the dataset by Wickramanayake & Bandara, (2016), the bus's fuel efficiency vs. speed curve follows a bell-shaped distribution. Therefore, the fuzzy values for the speed were selected as Low (L), Optimum (O), and High (H). The corresponding fuzzy values of acceleration are selected as Harsh Deceleration (HD), Acceptable (A), and Harsh Acceleration (HA) as the expert advice is to use gradual acceleration and breaking. We decided on these fuzzy values based on industrial norms (Verizon Telematics, n.d.) and the exploratory data analysis of the dataset (Wickramanayake & Bandara, 2016).

Finally, the fuzzy rules of the inference system (i.e., control actions) are derived as in Table II based on the fuzzy inputs and membership functions. Figure 8 shows the corresponding membership function of the fuzzy output, which are the actions to be taken by the driver.

Table II: Fuzzy rules.

| Speed | Acceleration | Control Action |
|-------|--------------|----------------|
| L | HD | Accelerate |
| L | HA | Accelerate Smoothly |
| O | HD | Keep the Speed |
| O | HA | Keep the Speed |
| H | HD | Break Smoothly |
| H | A | Break |
| H | HA | Break |

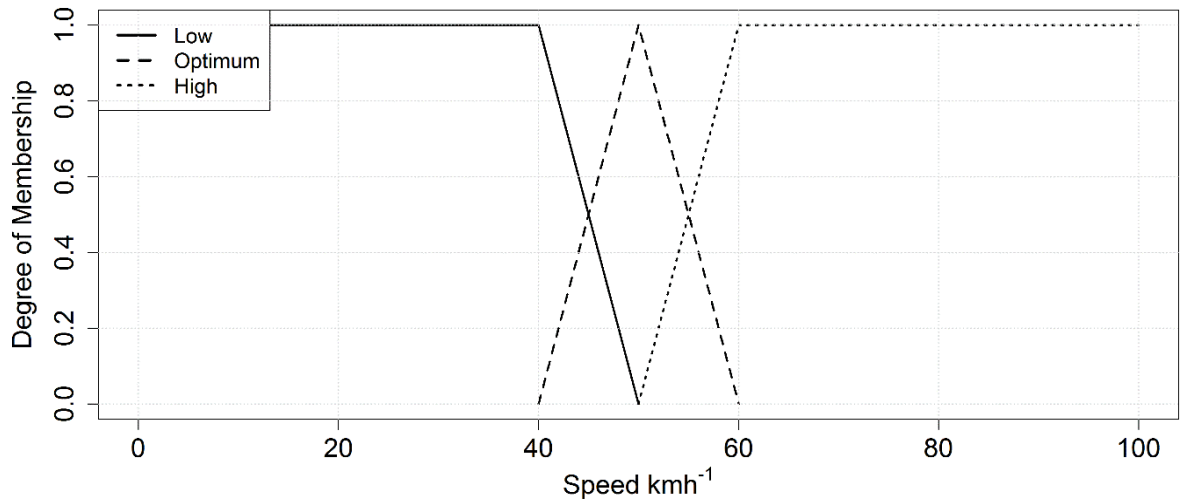

Figure 6: The membership function of speed.

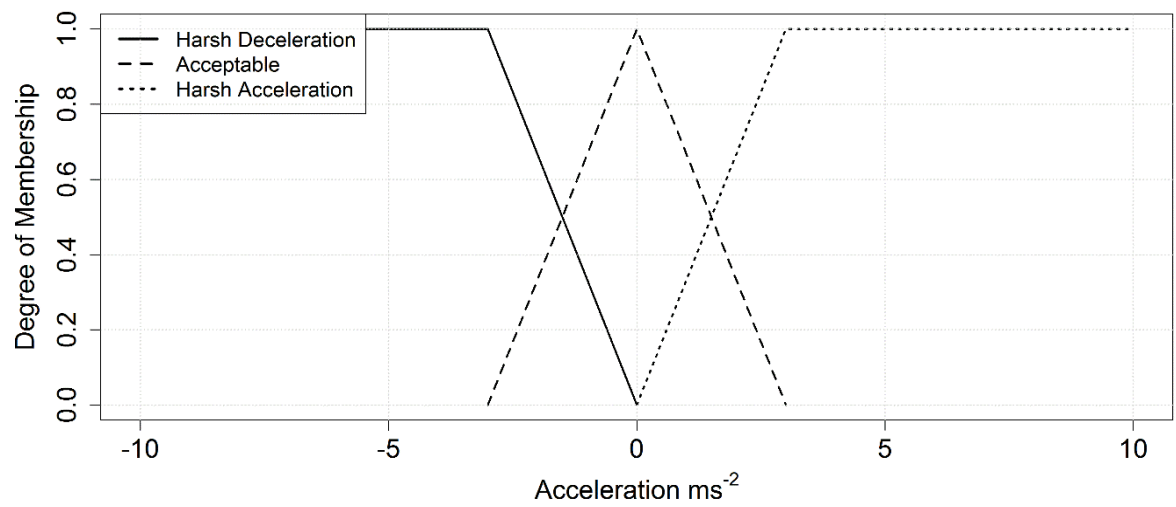

Figure 7: The membership function of acceleration.

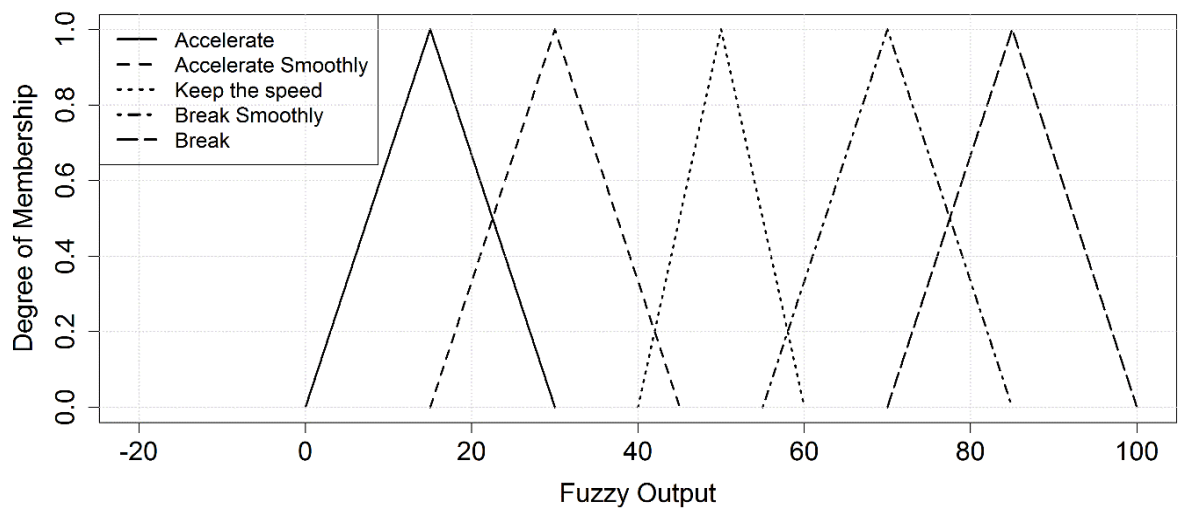

Figure 8: Fuzzy output membership function.

## V. PERFORMANCE ANALYSIS

The performance of the classification model was assessed using 10-fold cross-validation of the dataset mention in Section IV.A. To analyze the test results standard statistics were calculated. The results are given in Table III. Accuracy indicates the percentage of correctly classified test cases, whereas Kappa statistic shows the agreement of prediction with the true calls. While mean absolute error measures the average magnitude of errors of the prediction without considering the direction, root mean squared error gives the square root of the average squared error. Table III depicts that the random-forest based classifier has higher accuracy, precision, and recall while having a lower error. Higher precision and recall mean that most of the driving events identified as inefficient driving events are inefficient driving events and most inefficient driving events are correctly identified, respectively. Therefore, high precision and recall indicate that the feedback sent to the driver is not intrusive, as well as it is meaningful.

Table III: Statistics of results of the classification model.

| Selected Measure | Value |
|---|---|
| Accuracy | 85.16% |
| Kappa statistic | 0.7011 |
| Mean absolute error | 0.2082 |
| Root mean squared error | 0.3191 |
| Relative absolute error | 41.82% |
| Root relative squared error | 63.95% |
| Precision | 0.852 |
| Recall | 0.852 |

To test our solution's effectiveness in saving fuel, we carried out a simulation. We assessed to what extent the fuel efficiency of the bus can be improved if we would follow the historically best action for those detected inefficient driving events. Figure 9 shows both the actual fuel usage of the bus for a particular journey and the fuel usage when the historically best fuel efficiency replaces inefficient events under the same external conditions. It was identified that the estimated fuel usage, based on historically best fuel efficiency with driver feedback, is on average 16.36% higher than the actual fuel usage. This indicates an upper bound on the expected gain in fuel efficiency, as on any given day, the driver may not be able to drive at the best efficiency at every driving event. Nevertheless, aggregated saving over multiple days, routes, and vehicles could still be significant from the fleet owner's point of view, as the proposed solution is independent of the particular vehicle and route.

In Figure 9 there is one place (close to Latitude 80.7º) where actual fuel efficiency is better than the historically best case. This is because the classification model might have misclassified an efficient driving event as an inefficient driving event.

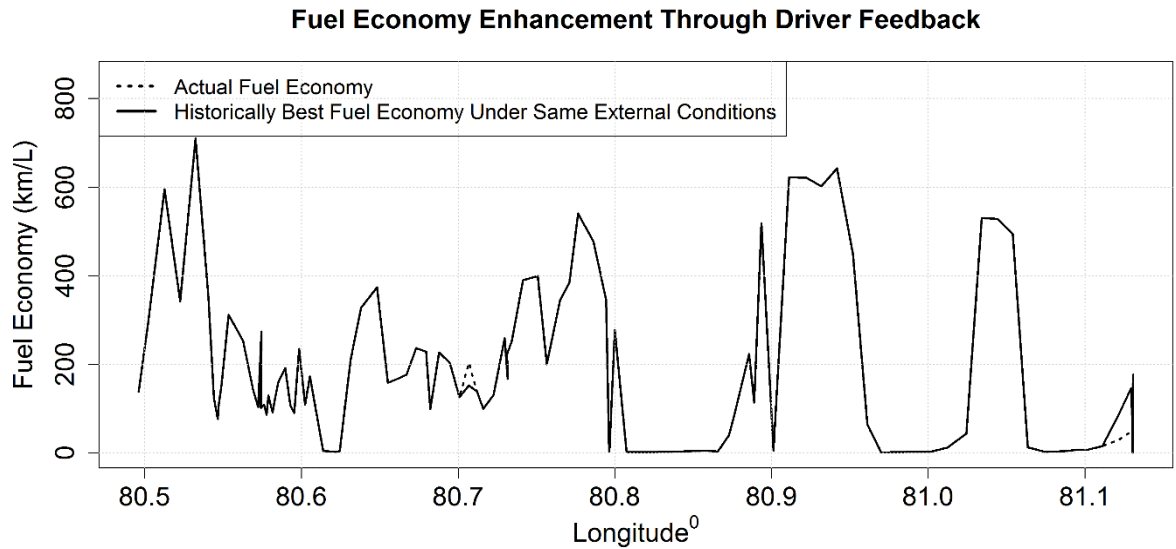

Figure 9: Actual fuel usage vs. adjusted fuel usage based on driver feedback for a selected journey.

## VI. Summary

We proposed a novel framework for promoting fuel-efficient driving behaviors among drivers via real-time driver monitoring and feedback. We demonstrated that by considering both the driver dependent parameters such as speed and acceleration, as well as external parameters such as weather, road traffic, and road topography, more accurate and useful feedback can be given to the driver. This was achieved by developing a random-forest based classification model to classify different driving behaviors as fuel efficient and inefficient. When a particular driving behavior is detected to be inefficient, a fuzzy-logic inference system was used to determine the corrective action to bring the vehicle back to a fuel-efficient state. Using a dataset from a long-distance bus, we demonstrated that the proposed solution could achieve significant fuel saving. As future work, we plan to integrate other driver-independent parameters such as the load of the vehicle, road type, and real-time traffic data. Being able to do the classification and fuzzy inference only on the smartphone is also of interest. This would eliminate the need to send data to the cloud in (near) real-time, saving both the bandwidth and power. While weather and traffic data need to be downloaded to the smartphone, it can be done less frequently.

## Acknowledgment

The authors are grateful for Nimbus Venture (Pvt) Ltd. for proving the dataset for the analysis.